\documentclass[conference]{IEEEtran}
\IEEEoverridecommandlockouts

\usepackage{cite}
\usepackage{amsmath,amssymb,amsfonts,amsthm}
\usepackage{graphicx}
\usepackage{textcomp}
\usepackage{booktabs}
\usepackage{tabularx}
\usepackage{array}
\usepackage{orcidlink}
\hypersetup{
  colorlinks=false,
  pdfborder={0 0 1},
  linkbordercolor={1 0 0},
  citebordercolor={0 1 0},
  urlbordercolor={0 0 1}
}
\newtheorem{theorem}{Theorem}

\providecommand{\doi}[1]{\href{https://doi.org/#1}{doi: #1}}

\def\BibTeX{{\rm B\kern-.05em{\sc i\kern-.025em b}\kern-.08em
    T\kern-.1667em\lower.7ex\hbox{E}\kern-.125emX}}

\begin{document}

\title{Optimal Dispatch of a Hydrogen-Colocated Renewable-Powered Desalination Plant}
\author{\IEEEauthorblockN{Guochen Zhao\orcidlink{0009-0008-8451-6368} and Ahmed S. Alahmed\orcidlink{0000-0002-4715-4379}}
\IEEEauthorblockA{\textit{Electrical Engineering Department, King Fahd University of Petroleum and Minerals}, Dhahran, KSA \\
\{g202424920, alahmad\}@kfupm.edu.sa}
\thanks{The work is supported in part by the Interdisciplinary Research Center of Smart Mobility \& Logistics at KFUPM under project INML2651.}
}

\maketitle

\begin{abstract}
This paper develops an analytical framework for profit-maximizing dispatch of water, electricity, and green hydrogen in a renewable-powered water desalination plant (WDP) combining thermal and reverse osmosis (RO) desalination. The optimal dispatch reveals that the schedules of the desalination units, electrolyzer, and grid interaction can all be characterized in closed form as functions of renewable generation. At low renewable output, the plant relies on thermal desalination and grid power while maintaining RO desalination and hydrogen production at their minimum setpoints. At high output, it reduces thermal desalination, increases RO desalination, and allocates surplus power and water to hydrogen production. For renewable output within a precomputed intermediate range, the WDP balances energy internally, allocating resources based on the relative marginal values of hydrogen and RO water. Simulations using real solar and plant data show that integrating RO and thermal desalination with renewable generation and hydrogen production achieves the highest daily profit of 75.945~k\$, exceeding the best configuration without hydrogen by 15.1\%.
\end{abstract}

\begin{IEEEkeywords}
Hydrogen production, optimal dispatch, reverse osmosis, thermal desalination, water-energy nexus.
\end{IEEEkeywords}

\section{Introduction}

The integration of green hydrogen production with water desalination plants (WDPs) has emerged as a promising option within the water--energy nexus. These colocated facilities physically and economically couple water extraction, hydrogen production, renewable generation, and grid exchange. In such systems, desalinated water is both a delivered product and an internal feedstock for electrolysis. Generated power must be allocated among thermal cogeneration, reverse osmosis (RO) desalination, electrolyzer operation, and grid exchange. Plant operators therefore face a coupled water--energy--hydrogen dispatch problem driven by market prices, renewable availability, and subsystem conversion efficiencies.

The coupling between water infrastructure and electric power systems has been widely studied \cite{Siddiqi2011,Oikonomou2020}. Meanwhile, RO has expanded rapidly and now dominates installed WDPs capacity, and it generally uses less energy than thermal desalination \cite{Eke2020}. Because desalination remains energy intensive, its integration with renewable energy has become an active research topic \cite{Ghaffour2015}.

Existing studies mainly consider RO-dominant frameworks, including standalone microgrids and coordinated scheduling platforms for grid interactions \cite{Guo2016,Mohammadi2019}. Analytical models have also characterized WDPs as flexible generator--load resources \cite{Alahmed2026}. However, these frameworks generally treat water as a terminal commodity and omit the internal dynamics introduced by colocated hydrogen production.

Electrolyzer scheduling under variable renewable supply has also been widely studied \cite{Wang2024RSER,Maluenda2023,Wang2025TSTE}. Closed-form policies have characterized the short-run profitability of renewable allocation between hydrogen production and wholesale power markets \cite{Li2025TechRxiv}. A numerical optimization study considers a coupled desalination--hydrogen system \cite{Liu2023JCP}. Analytical dispatch policies that capture the multi-carrier coupling of thermal desalination, RO, hydrogen production, and bidirectional grid participation remain absent from the literature.

This paper complements the work in \cite{Alahmed2026} by incorporating hydrogen production and market into the model and makes three main contributions.

First, we formulate a profit-maximizing scheduling model for a hydrogen-colocated desalination plant. The model captures the coupling among water, electricity, and hydrogen flows, including the water demand of the electrolyzer and the cogeneration behavior of the thermal desalination plant (TDP).

Second, we characterize the structure of the optimal dispatch policy and show that it follows a threshold-based form. The plant moves between import, net-zero, and export modes according to renewable availability. We also show that, in the net-zero mode, the internal dispatch depends on the relative marginal values of the RO desalination plant (RODP) and hydrogen production, which leads to hydrogen-priority and RO-priority cases.

Third, we validate the analytical results using local solar data. Through a comparative economic study, we show the value of operational flexibility and compare the daily profits of different plant configurations.

\section{Hydrogen-Colocated Renewable-Powered Desalination Plant}

\begin{figure}[t]
    \centering
    \includegraphics[width=\columnwidth]{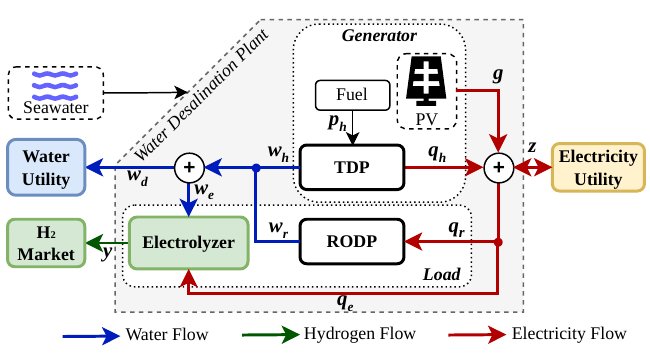}
    \vspace{-7mm}
    \caption{Schematic of the hydrogen-colocated renewable-powered WDP.}
    \label{fig:framework}
\end{figure}

The proposed \emph{profit-maximizing} desalination plant acts as a \emph{price-taking} agent colocated with renewable generation and an electrolyzer, simultaneously participating across electricity, water, and hydrogen markets (Fig.~\ref{fig:framework}). The plant integrates hybrid desalination technologies, specifically thermal and membrane-based methods. The TDP functions as a dual-market producer, yielding both water and electricity via cogeneration. Conversely, the RODP operates purely as an electrical load and a water producer. The electrolyzer bridges the multi-carrier system, acting as a load in both the electricity and water domains while producing hydrogen. Depending on the net electricity balance, the plant bidirectionally exchanges power with the utility grid.

\subsection{WDP Resources}

\subsubsection{TDP}
The TDP consumes fuel and jointly produces desalinated water and electricity. We adopt the linear conversion relationships $w_h=\alpha_h p_h$ \cite{AlNory2014} and $w_h=\eta_h q_h$ \cite{Santhosh2014}, where $\alpha_h,\eta_h,\beta_h \in \mathbb{R}_{++}$, $\alpha_h$ is the fuel-to-water conversion factor, $\eta_h$ is the water-to-electricity production ratio, and $\beta_h:=\alpha_h/\eta_h$ is the fuel-to-electricity conversion factor. We assume that the TDP fuel cost function $C_h(p_h)$ is strictly convex, continuously differentiable, and nondecreasing. The TDP water output satisfies $w_h\in[\underline{w}_h,\overline{w}_h]$, where $\underline{w}_h,\overline{w}_h \in \mathbb{R}_+$ are the minimum and maximum TDP water flowrates, respectively.

\subsubsection{RODP}
The RODP consumes electricity to produce desalinated water through membrane separation. We adopt a constant-specific-energy approximation for the RO unit. The approximation is $w_r=\alpha_r q_r$, where $\alpha_r\in\mathbb{R}_{++}$ is the water yield per unit electricity \cite{Mohammadi2019,Foroughian2026}. The RODP water output satisfies $w_r\in[\underline{w}_r,\overline{w}_r]$, where $\underline{w}_r,\overline{w}_r \in \mathbb{R}_+$ are the minimum and maximum RODP water flowrates, respectively.

\subsubsection{Electrolyzer}
The electrolyzer consumes electricity and desalinated water to produce hydrogen. We adopt a linear hydrogen production model for the electrolyzer \cite{Glenk2019}, extended to account for freshwater consumption \cite{Augustin2025}: $y=\gamma_e q_e$ and $w_e=\kappa_w y$, where $\gamma_e,\kappa_w \in \mathbb{R}_{++}$, $\gamma_e$ is the electricity-to-hydrogen conversion factor, and $\kappa_w$ is the water-use factor of hydrogen production. The electrolyzer power consumption is bounded by $q_e\in[\underline{q}_e,\overline{q}_e]$, where $\underline{q}_e,\overline{q}_e \in \mathbb{R}_+$ are the minimum and maximum power consumptions, respectively.

\subsubsection{Renewable Generation}
The aggregate renewable generation is denoted by $g \in \mathbb{R}_+$ (MW) and is treated as a known exogenous input over the scheduling interval.

\subsection{WDP Water, Electricity, and Hydrogen Payments}

The desalination plant is coupled through a water balance and an electricity balance. The total delivered water satisfies $w_d=w_h+w_r-w_e$ and $w_d\ge W_d$. Here, $W_d\in\mathbb{R}_+$ denotes the minimum required water delivery to the water utility. The plant's net electricity exchange with the utility grid is denoted by $z\in\mathbb{R}$. It satisfies $z=q_r+q_e-q_h-g$. The plant is in \emph{import mode} when $z>0$, in \emph{export mode} when $z<0$, and in \emph{net-zero mode} when $z=0$.

The plant receives revenue by selling water and hydrogen to the water and hydrogen markets. The corresponding revenues are $R_w(w_d)=\pi^w w_d$ and $R_h(y)=\pi^{H_2} y$, where $\pi^w,\pi^{H_2}\in\mathbb{R}_+$ are the water and hydrogen selling prices, respectively. Here, $P(z;g):=\pi^+[z]^+-\pi^-[z]^-$ denotes the WDP electricity payment. The electricity import and export prices are $\pi^+,\pi^-\in\mathbb{R}_+$, respectively, with $[x]^+:=\max\{x,0\}$ and $[x]^-:=\max\{-x,0\}$. This pricing structure is analogous to the net energy metering tariff, where the facility is billed based on its net electricity exchange \cite{Alahmed2023}.

\subsection{WDP Profit and Dispatch Problem}

The WDP's profit function is defined as
\begin{equation}
\begin{aligned}
\Pi(w_h,w_r,y,q_h,q_r,q_e;g)
&:=R_w(w_d)+R_h(y)\\
&\quad-P(z;g)-C_h(p_h).
\end{aligned}
\label{eq:ProfitFunction}
\end{equation}

Given the profit function in \eqref{eq:ProfitFunction}, the WDP's optimal dispatch is
\begin{align}
&(w_h^*,w_r^*,y^*,q_h^*,q_r^*,q_e^*) \notag\\
&\;:= \mathop{\mathrm{argmax}}_{\substack{w_h,w_r,y,q_h,q_r\\ q_e,p_h,w_d,z}}
\Pi(w_h,w_r,y,q_h,q_r,q_e;g) \label{eq:dispatch_problem}\\
\text{s.t.}\quad
& z=q_r+q_e-q_h-g, \notag\\
& w_d=w_h+w_r-w_e \ge W_d, \notag\\
& \underline{w}_h \le w_h \le \overline{w}_h, \notag\\
& \underline{w}_r \le w_r \le \overline{w}_r, \notag\\
& \underline{q}_e \le q_e \le \overline{q}_e, \notag\\
& w_h=\alpha_h p_h=\eta_h q_h, \notag\\
& w_r=\alpha_r q_r, \notag\\
& y=\gamma_e q_e,\qquad w_e=\kappa_w y.
\notag
\end{align}
We focus on operating conditions under which the delivered-water constraint $w_d\ge W_d$ is nonbinding at the optimum. The constraint therefore does not affect the optimal threshold structure. Given $\pi^+ \ge \pi^-$, the program in \eqref{eq:dispatch_problem} is strictly concave in $w_h$ and concave in $w_r$ and $y$, and therefore has a global optimum.

\section{Optimal Dispatch Policy}
\label{sec:OptDis}

We show that the optimal dispatch can be described by a set of renewable-generation thresholds. Depending on the level of renewable output, the plant operates in \emph{import mode} ($z>0$), \emph{net-zero mode} ($z=0$), or \emph{export mode} ($z<0$). In each mode, the plant adjusts its internal allocation to maximize profit.

\subsection{Threshold-Based Dispatch}

To study the internal allocation priority, we define the effective marginal values of electricity directed to RO desalination and electrolysis as $\rho_r:=\alpha_r\pi^w$ and $\rho_e:=\gamma_e\pi^{H_2}-\gamma_e\kappa_w\pi^w$, respectively. Their gap, $\Delta\rho:=\rho_e-\rho_r$, determines whether hydrogen production or RO desalination has priority in the net-zero mode.

During the import and export states, the grid sets the marginal price, and the TDP optimizes output based on standard stationarity conditions
\begin{equation*}
C_h'(p_h^*)=\beta_h\pi^+ + \alpha_h\pi^w
\quad \text{or} \quad
C_h'(p_h^*)=\beta_h\pi^- + \alpha_h\pi^w.
\end{equation*}
These yield the following TDP water flows in the import and export modes, respectively
\begin{equation*}
\begin{aligned}
w_h^{\mbox{\tiny IM}}
&=\left[\alpha_h D_h\!\left(\alpha_h\pi^w+\beta_h\pi^+\right)\right]_{[\underline{w}_h,\overline{w}_h]},\\
w_h^{\mbox{\tiny EX}}
&=\left[\alpha_h D_h\!\left(\alpha_h\pi^w+\beta_h\pi^-\right)\right]_{[\underline{w}_h,\overline{w}_h]},
\end{aligned}
\end{equation*}
where $D_h:=(C_h')^{-1}$ is the inverse marginal fuel-cost function, and $[x]_{[a,b]}:=\min\{\max\{x,a\},b\}$ denotes the projection of $x$ onto the interval $[a,b]$.

For operations within the net-zero mode, we define the internal shadow-price function
\begin{equation*}
q_h(\rho):=
\frac{1}{\eta_h}
\left[\alpha_h D_h\!\left(\alpha_h\pi^w+\beta_h\rho\right)\right]_{[\underline{w}_h,\overline{w}_h]}.
\end{equation*}
We also define the RO unit's power bounds
{
\begin{equation*}
\underline{q}_r:=\frac{\underline{w}_r}{\alpha_r},\qquad
\overline{q}_r:=\frac{\overline{w}_r}{\alpha_r}.
\end{equation*}
}

\noindent The dispatch policy is formalized in the following theorem for
$\rho_r,\rho_e\in(\pi^-,\pi^+)$. Cases violating this condition are
treated in Appendix~\ref{app:other_orders}.

\begin{theorem}[WDP Optimal dispatch]
\label{thm:optimal_dispatch}
Assume the marginal values are distinct and strictly interior to the grid prices. Equivalently, $\rho_r,\rho_e\in(\pi^-,\pi^+)$ and $\rho_r\ne\rho_e$. The optimal dispatch has six thresholds defining seven dispatch regions. First, the outer boundaries separating grid interactions are defined as
{
\begin{equation*}
\Gamma^{\mbox{\tiny IM}}:=\underline{q}_r+\underline{q}_e-q_h(\pi^+),\qquad
\Gamma^{\mbox{\tiny EX}}:=\overline{q}_r+\overline{q}_e-q_h(\pi^-),
\end{equation*}
}
categorizing the net electricity exchange into three modes
\begin{equation*}
z^*(g)=
\begin{cases}
>0, & g<\Gamma^{\mbox{\tiny IM}},\\
=0, & \Gamma^{\mbox{\tiny IM}}\le g\le \Gamma^{\mbox{\tiny EX}},\\
<0, & g>\Gamma^{\mbox{\tiny EX}}.
\end{cases}
\end{equation*}

Second, if the market satisfies $\pi^-<\rho_r<\rho_e<\pi^+$, the plant follows a hydrogen-priority dispatch. The net-zero thresholds are
{
\begin{equation*}
\begin{aligned}
\Gamma_1 &:= \underline{q}_r+\underline{q}_e-q_h(\rho_e), \quad
\Gamma_2 := \underline{q}_r+\overline{q}_e-q_h(\rho_e),\\
\Gamma_3 &:= \underline{q}_r+\overline{q}_e-q_h(\rho_r), \quad
\Gamma_4 := \overline{q}_r+\overline{q}_e-q_h(\rho_r).
\end{aligned}
\end{equation*}
}
Under this regime, the closed-form schedules are
{
\begin{equation}
w_h^*(g)=
\begin{cases}
w_h^{\mbox{\tiny IM}}, & g<\Gamma^{\mbox{\tiny IM}},\\
\eta_h\!\left(\underline{q}_r+\underline{q}_e-g\right), & g\in[\Gamma^{\mbox{\tiny IM}},\Gamma_1),\\
\eta_h q_h(\rho_e), & g\in[\Gamma_1,\Gamma_2),\\
\eta_h\!\left(\underline{q}_r+\overline{q}_e-g\right), & g\in[\Gamma_2,\Gamma_3),\\
\eta_h q_h(\rho_r), & g\in[\Gamma_3,\Gamma_4),\\
\eta_h\!\left(\overline{q}_r+\overline{q}_e-g\right), & g\in[\Gamma_4,\Gamma^{\mbox{\tiny EX}}],\\
w_h^{\mbox{\tiny EX}}, & g>\Gamma^{\mbox{\tiny EX}},
\end{cases}
\label{eq:wh_star_h2}
\end{equation}
}
\begin{equation}
w_r^*(g)=
\begin{cases}
\underline{w}_r, & g<\Gamma_3,\\
\alpha_r\!\left(q_h(\rho_r)+g-\overline{q}_e\right), & g\in[\Gamma_3,\Gamma_4],\\
\overline{w}_r, & g\ge\Gamma_4,
\end{cases}
\label{eq:wr_star_h2}
\end{equation}
{
\begin{equation}
\begin{aligned}
y^*(g)&=\gamma_e q_e^*(g),\\
q_e^*(g)&=
\begin{cases}
\underline{q}_e, & g<\Gamma_1,\\
q_h(\rho_e)-\underline{q}_r+g, & g\in[\Gamma_1,\Gamma_2),\\
\overline{q}_e, & g\ge \Gamma_2.
\end{cases}
\end{aligned}
\label{eq:y_star_h2}
\end{equation}
}

\begin{figure}[t]
    \centering
    \includegraphics[width=\columnwidth]{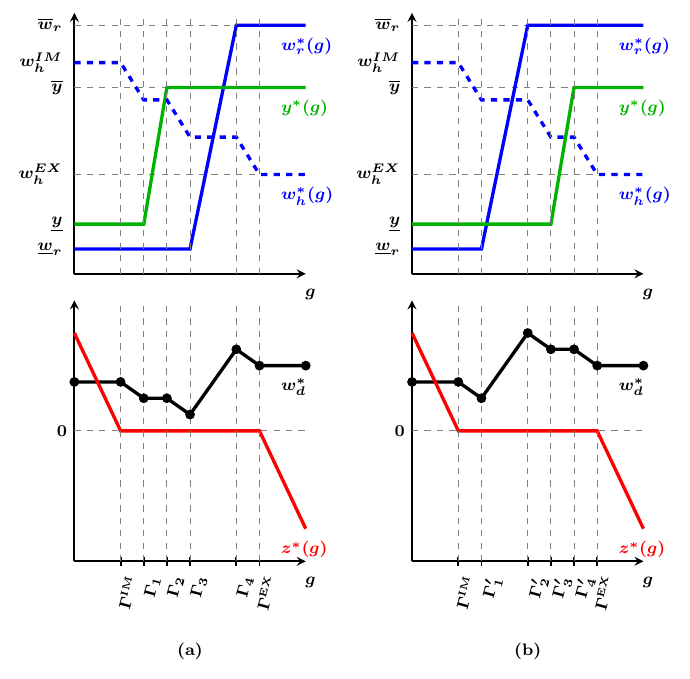}
    \vspace{-4mm}
    \vspace{-5mm}
    \caption{Optimal WDP dispatch under two priority cases: (a) hydrogen-priority case with $\rho_e > \rho_r$; (b) RO-priority case with $\rho_r > \rho_e$.}
    \label{fig:dual_flex}
\end{figure}

Third, if instead $\pi^-<\rho_e<\rho_r<\pi^+$, the plant shifts to RO-priority dispatch. The modified internal boundaries are
{
\begin{equation*}
\begin{aligned}
\Gamma_1' &:= \underline{q}_r+\underline{q}_e-q_h(\rho_r), \quad
\Gamma_2' := \overline{q}_r+\underline{q}_e-q_h(\rho_r),\\
\Gamma_3' &:= \overline{q}_r+\underline{q}_e-q_h(\rho_e), \quad
\Gamma_4' := \overline{q}_r+\overline{q}_e-q_h(\rho_e).
\end{aligned}
\end{equation*}
}
The corresponding flow allocations are
{
\begin{equation}
w_h^*(g)=
\begin{cases}
w_h^{\mbox{\tiny IM}}, & g<\Gamma^{\mbox{\tiny IM}},\\
\eta_h\!\left(\underline{q}_r+\underline{q}_e-g\right), & g\in[\Gamma^{\mbox{\tiny IM}},\Gamma_1'),\\
\eta_h q_h(\rho_r), & g\in[\Gamma_1',\Gamma_2'),\\
\eta_h\!\left(\overline{q}_r+\underline{q}_e-g\right), & g\in[\Gamma_2',\Gamma_3'),\\
\eta_h q_h(\rho_e), & g\in[\Gamma_3',\Gamma_4'),\\
\eta_h\!\left(\overline{q}_r+\overline{q}_e-g\right), & g\in[\Gamma_4',\Gamma^{\mbox{\tiny EX}}],\\
w_h^{\mbox{\tiny EX}}, & g>\Gamma^{\mbox{\tiny EX}},
\end{cases}
\label{eq:wh_star_water}
\end{equation}
}
{
\begin{equation}
w_r^*(g)=
\begin{cases}
\underline{w}_r, & g<\Gamma_1',\\
\alpha_r\!\left(q_h(\rho_r)+g-\underline{q}_e\right), & g\in[\Gamma_1',\Gamma_2'),\\
\overline{w}_r, & g\ge \Gamma_2',
\end{cases}
\label{eq:wr_star_water}
\end{equation}
}
{
\begin{align}
y^*(g)&=\gamma_e q_e^*(g),\nonumber\\
q_e^*(g)&=
\begin{cases}
\underline{q}_e, & g<\Gamma_3',\\
q_h(\rho_e)-\overline{q}_r+g, & g\in[\Gamma_3',\Gamma_4'),\\
\overline{q}_e, & g\ge \Gamma_4'.
\end{cases}
\label{eq:y_star_water}
\end{align}
}
\end{theorem}

The dispatch schedules in \eqref{eq:wh_star_h2}--\eqref{eq:y_star_water} are illustrated in Fig.~\ref{fig:dual_flex}. In Fig.~\ref{fig:dual_flex}(a), when $\rho_e>\rho_r$, additional renewable power in the net-zero region is first allocated to the electrolyzer until it reaches its upper limit, and then to the RO unit. In Fig.~\ref{fig:dual_flex}(b), when $\rho_r>\rho_e$, the plant gives priority to RO production because the marginal value of water is higher than that of hydrogen. The ordering of $\rho_r$ and $\rho_e$ affects the dispatch only in the net-zero mode. The proof is provided in Appendix~\ref{app:proof_thm1}.

These thresholds define the three operating modes. The plant imports for $g<\Gamma^{\mbox{\tiny IM}}$ and exports for $g>\Gamma^{\mbox{\tiny EX}}$. It is in net-zero mode for $\Gamma^{\mbox{\tiny IM}}\le g\le \Gamma^{\mbox{\tiny EX}}$.

\section{Numerical Results}

\subsection{Parameter Setting and Dispatch Validation}

We validate the analytical framework using a 2019 solar generation profile for Dhahran, Saudi Arabia, derived from MERRA-2 data \cite{Gelaro2017} and scaled to a 115 MW array. For the numerical study, the import price is $\pi^+=120\$/\mathrm{MWh}$, and the export price is $\pi^-=40\$/\mathrm{MWh}$. We use representative conversion parameters. Here, $\alpha_r=166.67\mathrm{m}^3/\mathrm{MWh}$ is the RO water yield. For the electrolyzer, $\gamma_e=18.2\mathrm{kg}/\mathrm{MWh}$ and $\kappa_w=0.010\mathrm{m}^3/\mathrm{kg}$. For the TDP, $\alpha_h=4.0\mathrm{m}^3/\mathrm{MBTU}$ and $\eta_h=80.0\mathrm{m}^3/\mathrm{MWh}$. The desalination parameters follow the benchmark setting in \cite{Alahmed2026}, while the electrolyzer parameters are consistent with the ranges reviewed in \cite{Augustin2025}. The TDP fuel cost is modeled as $C_h(p_h)=a p_h^2+b p_h+c$,
with $a=0.008\$/\mathrm{MBTU}^2$, $b=2.0\$/\mathrm{MBTU}$, and $c=0$ \cite{Santhosh2014}.
The minimum operating limits are $\underline{q}_e=\underline{w}_r=\underline{w}_h=0$. The electrolyzer upper limit is $\overline{q}_e=30\mathrm{MW}$. The RODP and TDP upper limits are $\overline{w}_r=8333\mathrm{m}^3/\mathrm{h}$ and $\overline{w}_h=3000\mathrm{m}^3/\mathrm{h}$, respectively.

We compare two price scenarios. In Scenario A, the water and hydrogen prices $(\pi^w,\pi^{H_2})$ are 0.50 and 5.50, respectively; under these prices, hydrogen has priority over RO. In Scenario B, they are 0.65 and 4.50, respectively; under these prices, RO has priority over hydrogen.

\begin{figure}[t]
    \centering
    \includegraphics[width=\columnwidth]{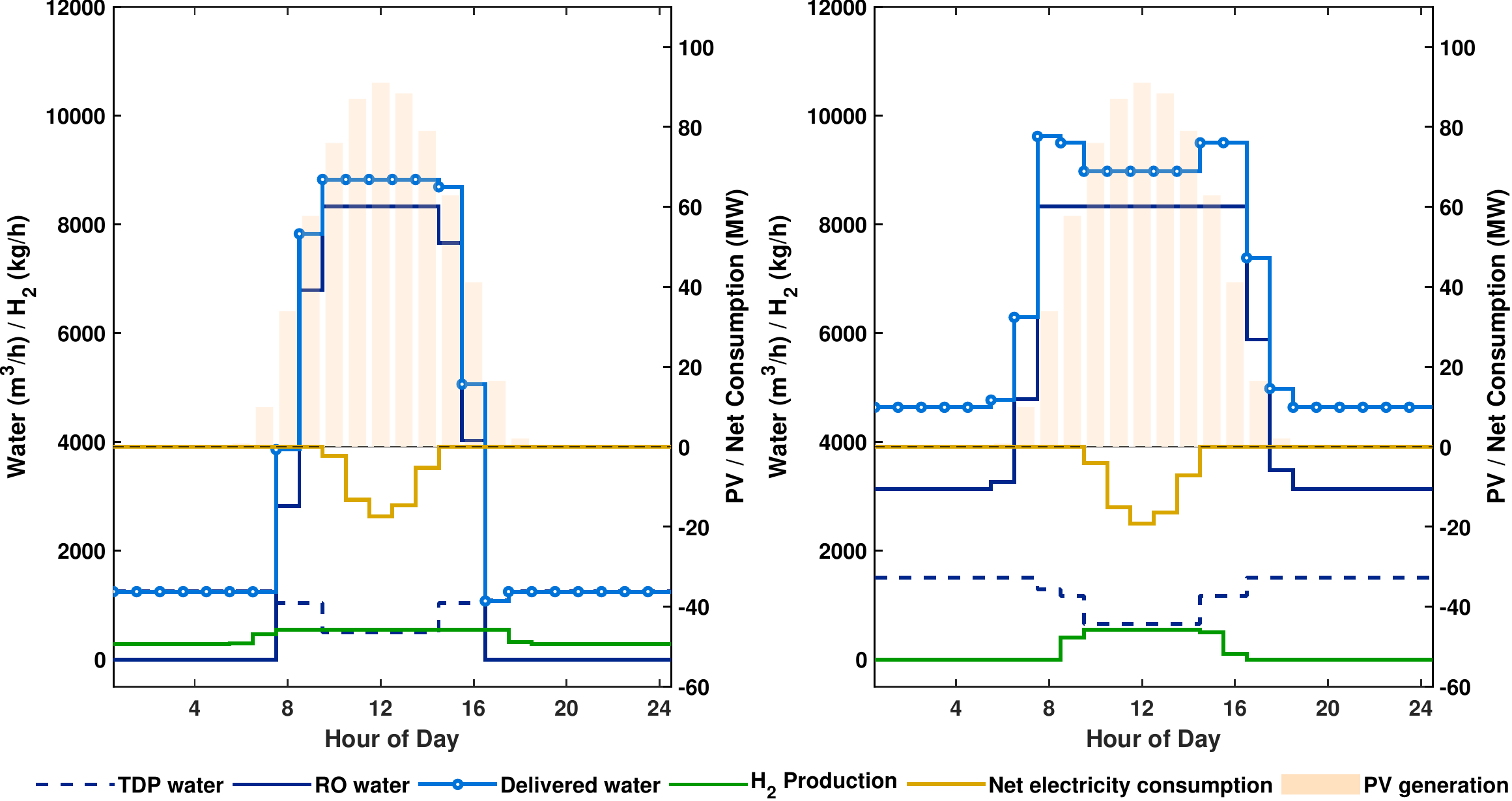}
    \caption{WDP 24-hour dispatch under two price cases: (left) hydrogen-priority dispatch with $\rho_e>\rho_r$; (right) RO-priority dispatch with $\rho_e<\rho_r$.}
    \label{fig:24h_dispatch}
\end{figure}

Fig.~\ref{fig:24h_dispatch} illustrates the two dispatch scenarios. In the left panel, as solar generation increases in the morning, the electrolyzer absorbs the additional renewable power first, and the RO unit starts to increase only after the electrolyzer reaches its limit. In the right panel, the order is reversed because water has a higher marginal value than hydrogen, so the RO unit increases first and the electrolyzer starts later. In both cases, the plant operates in net-zero mode at low and moderate solar output and moves to grid export near peak solar output; the import mode is absent because $\Gamma^{\mbox{\tiny IM}}<0$.

\subsection{Daily Profit Across Different Resource Compositions}

\begin{table}[t]
\caption{Daily WDP profit under different technology compositions}
\label{tab:daily_results}
\centering
\scriptsize
\setlength{\tabcolsep}{2.4pt}
\renewcommand{\arraystretch}{1.08}
\begin{tabularx}{\columnwidth}{@{}>{\raggedright\arraybackslash}p{0.25\columnwidth}*{5}{>{\centering\arraybackslash}X}@{}}
\toprule
\textbf{Metric} & \textbf{RO+PV} & \textbf{TDP+PV} & \textbf{\shortstack{RO+PV\\+H$_2$}} & \textbf{\shortstack{RO+TDP\\+PV}} & \textbf{Full} \\
\midrule
Water rev. (k\$)      & 43.539 &  7.935 & 43.521 & 83.219 & 84.218 \\
H$_2$ rev. (k\$)      &  0.000 &  0.000 & 15.078 &  0.000 & 18.856 \\
Elec. net rev. (k\$)  &  7.688 & 32.760 &  1.060 &  9.700 &  2.299 \\
Fuel cost (k\$)       &  0.000 & 10.868 &  0.000 & 26.922 & 29.428 \\
\midrule
Profit (k\$)          & 51.226 & 29.828 & 59.659 & 65.998 & 75.945 \\
\bottomrule
\end{tabularx}
\end{table}

To evaluate the economic value of system integration, Table~\ref{tab:daily_results} compares five plant configurations over an average daily cycle. The comparison uses $\pi^w=0.575\$/\mathrm{m}^3$ and $\pi^{H_2}=5.0\$/\mathrm{kg}$. The results show that RO+PV performs better than TDP+PV among the two single-desalination configurations, mainly because it avoids fuel cost. Although TDP+PV earns more from electricity sales, much of that benefit is offset by thermal fuel consumption. Adding hydrogen production to RO+PV increases the daily profit from 51.226 k\$ to 59.659 k\$, but it still does not outperform the RO+TDP+PV configuration, whose profit reaches 65.998 k\$. Thus, under the current price and capacity setting, adding green-hydrogen production alone is not the most profitable option. The highest profit, 75.945 k\$, is achieved by the full RO+TDP+PV+H$_2$ system, indicating that hydrogen production has a higher value when coordinated with hybrid desalination.

\section{Conclusion}

We presented an analytical dispatch framework for a renewable-powered hybrid desalination plant with colocated hydrogen production. We showed that the profit-maximizing dispatch follows a threshold-based structure with import, net-zero, and export modes. In the net-zero mode, the internal allocation depends on the relative marginal values of hydrogen production and RO desalination, which leads to hydrogen-priority and RO-priority operating patterns. Numerical results based on local solar data supported the analytical results. The daily economic comparison showed that hydrogen production can improve profit, but its value is higher when it is combined with a hybrid desalination plant rather than with a single desalination technology. Among the cases studied, the full RO+TDP+PV+H$_2$ configuration achieved the highest profit.

Several aspects are omitted from this short-run analysis. The conversion functions may exhibit highly nonlinear behavior under different operating conditions, which is not captured in this paper. Large-scale water and hydrogen storage and distribution costs are also outside the short-run dispatch model considered here. Incorporating these factors into long-term planning models remains a direction for future work.

\appendices

\section{Proof of Theorem~\ref{thm:optimal_dispatch}}
\label{app:proof_thm1}

Using
\begin{align*}
w_h &= \eta_h q_h, & w_r &= \alpha_r q_r, & y &= \gamma_e q_e, \\
w_e &= \kappa_w\gamma_e q_e, & p_h &= \frac{q_h}{\beta_h},
\end{align*}
and recalling
\[
\rho_r=\alpha_r\pi^w,\qquad
\rho_e=\gamma_e\pi^{H_2}-\gamma_e\kappa_w\pi^w,
\]
the dispatch problem can be equivalently written, up to constants
independent of the decision variables, as
\begin{subequations}
\label{eq:app_dispatch_q}
\begin{align}
\max_{q_h,q_r,q_e}\quad
& \eta_h\pi^w q_h+\rho_r q_r+\rho_e q_e
-P(z;g)-C_h\!\left(\frac{q_h}{\beta_h}\right), \nonumber\\
\text{s.t.}\quad
& z=q_r+q_e-q_h-g, \label{eq:app_balance}\\
& q_h \in [\underline{q}_h,\overline{q}_h], \label{eq:app_qh_bound}\\
& q_r \in [\underline{q}_r,\overline{q}_r], \label{eq:app_qr_bound}\\
& q_e \in [\underline{q}_e,\overline{q}_e], \label{eq:app_qe_bound}
\end{align}
\end{subequations}
where
\[
\underline{q}_h:=\frac{\underline{w}_h}{\eta_h},\qquad
\overline{q}_h:=\frac{\overline{w}_h}{\eta_h},\qquad
\underline{q}_r:=\frac{\underline{w}_r}{\alpha_r},\qquad
\overline{q}_r:=\frac{\overline{w}_r}{\alpha_r}.
\]
Under the nonbinding-water-constraint assumption $w_d\geq W_d$, the delivered-water constraint can be omitted.

Equivalently, we consider the convex minimization problem
\begin{equation}
\min_{q_h,q_r,q_e}\;
C_h\!\left(\frac{q_h}{\beta_h}\right)
-\eta_h\pi^w q_h-\rho_r q_r-\rho_e q_e
+P(z;g),
\label{eq:app_reduced_min}
\end{equation}
subject to \eqref{eq:app_balance}--\eqref{eq:app_qe_bound}.

We analyze \eqref{eq:app_reduced_min} in three modes:
\[
\mbox{\small IM}: z>0,\qquad
\mbox{\small NZ}: z=0,\qquad
\mbox{\small EX}: z<0.
\]
In each mode, the objective is convex and the constraints are affine or box
constraints. Therefore, the KKT conditions are necessary and sufficient for
optimality.

For any effective marginal electricity value $\xi$, recall the TDP dispatch function
\begin{equation*}
q_h(\xi):=
\frac{1}{\eta_h}
\left[
\alpha_hD_h\!\left(\alpha_h\pi^w+\beta_h\xi\right)
\right]_{[\underline{w}_h,\overline{w}_h]},
\end{equation*}
where $D_h=(C_h')^{-1}$ is the inverse marginal fuel-cost function.

\subsubsection*{1) Import and export modes}

In the import mode, $z>0$, and $P(z;g)=\pi^+z$. Hence
\eqref{eq:app_reduced_min} becomes
\[
C_h\!\left(\frac{q_h}{\beta_h}\right)
-(\eta_h\pi^w+\pi^+)q_h
+(\pi^+-\rho_r)q_r
+(\pi^+-\rho_e)q_e
-\pi^+g.
\]
Since $\rho_r,\rho_e<\pi^+$, increasing either $q_r$ or $q_e$ increases this
minimization objective. Hence, the KKT conditions over the box constraints
give
\[
q_r^{\mbox{\tiny IM}}=\underline{q}_r,\qquad
q_e^{\mbox{\tiny IM}}=\underline{q}_e.
\]
For the TDP, the stationarity condition with respect to $q_h$ is
\[
0=
\frac{1}{\beta_h}C_h'\!\left(\frac{q_h}{\beta_h}\right)
-\eta_h\pi^w-\pi^+.
\]
Equivalently, using $p_h=q_h/\beta_h$ and $\alpha_h=\beta_h\eta_h$,
\[
C_h'(p_h)=\alpha_h\pi^w+\beta_h\pi^+.
\]
Including the TDP bounds through complementary slackness yields
\[
q_h^{\mbox{\tiny IM}}=q_h(\pi^+)
=
\frac{w_h^{\mbox{\tiny IM}}}{\eta_h},
\]
where
\[
w_h^{\mbox{\tiny IM}}
=
\left[
\alpha_hD_h\!\left(\alpha_h\pi^w+\beta_h\pi^+\right)
\right]_{[\underline{w}_h,\overline{w}_h]}.
\]
Therefore,
\[
z^{\mbox{\tiny IM}}(g)
=
\underline{q}_r+\underline{q}_e-q_h(\pi^+)-g,
\]
and the import solution is valid when
\[
g<\Gamma^{\mbox{\tiny IM}}
:=
\underline{q}_r+\underline{q}_e-q_h(\pi^+)
=
\underline{q}_r+\underline{q}_e-\frac{w_h^{\mbox{\tiny IM}}}{\eta_h}.
\]

In the export mode, $z<0$, and $P(z;g)=\pi^-z$. Hence
\eqref{eq:app_reduced_min} becomes
\[
C_h\!\left(\frac{q_h}{\beta_h}\right)
-(\eta_h\pi^w+\pi^-)q_h
+(\pi^--\rho_r)q_r
+(\pi^--\rho_e)q_e
-\pi^-g.
\]
Since $\pi^-<\rho_r,\rho_e$, increasing either $q_r$ or $q_e$ decreases this
minimization objective. Hence, the KKT conditions over the box constraints
give
\[
q_r^{\mbox{\tiny EX}}=\overline{q}_r,\qquad
q_e^{\mbox{\tiny EX}}=\overline{q}_e.
\]
For the TDP, the stationarity condition with respect to $q_h$ is
\[
0=
\frac{1}{\beta_h}C_h'\!\left(\frac{q_h}{\beta_h}\right)
-\eta_h\pi^w-\pi^-.
\]
Equivalently,
\[
C_h'(p_h)=\alpha_h\pi^w+\beta_h\pi^-.
\]
Including the TDP bounds through complementary slackness yields
\[
q_h^{\mbox{\tiny EX}}=q_h(\pi^-)
=
\frac{w_h^{\mbox{\tiny EX}}}{\eta_h},
\]
where
\[
w_h^{\mbox{\tiny EX}}
=
\left[
\alpha_hD_h\!\left(\alpha_h\pi^w+\beta_h\pi^-\right)
\right]_{[\underline{w}_h,\overline{w}_h]}.
\]
Therefore,
\[
z^{\mbox{\tiny EX}}(g)
=
\overline{q}_r+\overline{q}_e-q_h(\pi^-)-g,
\]
and the export solution is valid when
\[
g>\Gamma^{\mbox{\tiny EX}}
:=
\overline{q}_r+\overline{q}_e-q_h(\pi^-)
=
\overline{q}_r+\overline{q}_e-\frac{w_h^{\mbox{\tiny EX}}}{\eta_h}.
\]

\subsubsection*{2) Net-zero mode}

In the net-zero mode,
\[
z=0,\qquad q_r+q_e-q_h-g=0.
\]
The net-zero mode problem becomes
\begin{equation*}
\min_{q_h,q_r,q_e}\;
C_h\!\left(\frac{q_h}{\beta_h}\right)
-\eta_h\pi^wq_h-\rho_rq_r-\rho_eq_e
\end{equation*}
subject to the net-zero balance and the box constraints.

Let $\mu$ be the Lagrange multiplier of the net-zero balance
$q_r+q_e-q_h-g=0$. It represents the internal marginal value of electricity
in the net-zero mode. The Lagrangian is
\[
\begin{aligned}
\mathcal L^{\mbox{\tiny NZ}}
=&
C_h\!\left(\frac{q_h}{\beta_h}\right)
-\eta_h\pi^wq_h
-\rho_rq_r
-\rho_eq_e\\
&+\mu(q_r+q_e-q_h-g)\\
&+\lambda_h^-(\underline{q}_h-q_h)
+\lambda_h^+(q_h-\overline{q}_h)\\
&+\lambda_r^-(\underline{q}_r-q_r)
+\lambda_r^+(q_r-\overline{q}_r)\\
&+\lambda_e^-(\underline{q}_e-q_e)
+\lambda_e^+(q_e-\overline{q}_e),
\end{aligned}
\]
where all inequality multipliers are nonnegative. The stationarity
conditions are
\begin{align}
0&=
\frac{1}{\beta_h}C_h'\!\left(\frac{q_h^\ast}{\beta_h}\right)
-\eta_h\pi^w-\mu-\lambda_h^-+\lambda_h^+,
\label{eq:app_nz_stat_h}\\
0&=
-\rho_r+\mu-\lambda_r^-+\lambda_r^+,
\label{eq:app_nz_stat_r}\\
0&=
-\rho_e+\mu-\lambda_e^-+\lambda_e^+.
\label{eq:app_nz_stat_e}
\end{align}
Together with complementary slackness, \eqref{eq:app_nz_stat_h} gives
\[
q_h^\ast=q_h(\mu).
\]

We next derive the dispatch of the RO unit from
\eqref{eq:app_nz_stat_r}. The complementary slackness conditions for
$q_r$ are
\[
\lambda_r^-(\underline{q}_r-q_r)=0,\qquad
\lambda_r^+(q_r-\overline{q}_r)=0,
\qquad
\lambda_r^-,\lambda_r^+\ge 0.
\]
If $q_r$ is strictly inside its bounds, then
$\lambda_r^-=\lambda_r^+=0$, and \eqref{eq:app_nz_stat_r} implies
$\mu=\rho_r$. Hence, the RO unit can be marginal only when
$\mu=\rho_r$. If $\mu>\rho_r$, then the stationarity condition can be
satisfied only with the lower bound active, so $q_r=\underline{q}_r$.
If $\mu<\rho_r$, it can be satisfied only with the upper bound active, so
$q_r=\overline{q}_r$. Therefore,
\begin{equation}
q_r^\ast(\mu)=
\begin{cases}
\overline{q}_r, & \mu<\rho_r,\\
[\underline{q}_r,\overline{q}_r], & \mu=\rho_r,\\
\underline{q}_r, & \mu>\rho_r.
\end{cases}
\label{eq:app_qr_mu}
\end{equation}

The same argument applied to \eqref{eq:app_nz_stat_e}, together with the
complementary slackness conditions for $q_e$, gives
\begin{equation}
q_e^\ast(\mu)=
\begin{cases}
\overline{q}_e, & \mu<\rho_e,\\
[\underline{q}_e,\overline{q}_e], & \mu=\rho_e,\\
\underline{q}_e, & \mu>\rho_e.
\end{cases}
\label{eq:app_qe_mu}
\end{equation}
At $z=0$, a small increase in net consumption is priced at $\pi^+$, while a
small decrease is valued at $\pi^-$. Therefore, the net-zero shadow price
must satisfy
\[
\mu\in[\pi^-,\pi^+].
\]
Moreover, since $C_h'(\cdot)$ and $D_h(\cdot)$ are nondecreasing and
projection preserves monotonicity, $q_h(\mu)$ is nondecreasing in $\mu$.

\subsubsection*{3) Hydrogen-priority case}

Suppose
\[
\pi^-<\rho_r<\rho_e<\pi^+.
\]
As the net-zero shadow price moves from $\pi^+$ toward $\pi^-$, it reaches
$\rho_e$ before $\rho_r$. Hence, the electrolyzer becomes marginal before
the RO unit.

The internal thresholds are obtained from the net-zero balance
\[
g=q_r+q_e-q_h.
\]
They correspond to the points where the electrolyzer or the RO unit starts
or stops being marginal. In the hydrogen-priority case, these points are
\begin{align*}
\Gamma_1 &:= \underline{q}_r+\underline{q}_e-q_h(\rho_e),
&
\Gamma_2 &:= \underline{q}_r+\overline{q}_e-q_h(\rho_e),\\
\Gamma_3 &:= \underline{q}_r+\overline{q}_e-q_h(\rho_r),
&
\Gamma_4 &:= \overline{q}_r+\overline{q}_e-q_h(\rho_r).
\end{align*}
Here, $\Gamma_1$ and $\Gamma_2$ are obtained when $\mu=\rho_e$ and the
electrolyzer moves from $\underline{q}_e$ to $\overline{q}_e$, while the RO
unit remains at $\underline{q}_r$. Similarly, $\Gamma_3$ and $\Gamma_4$ are
obtained when $\mu=\rho_r$ and the RO unit moves from $\underline{q}_r$ to
$\overline{q}_r$, while the electrolyzer remains at $\overline{q}_e$.

Since $q_h(\mu)$ is nondecreasing,
\[
q_h(\pi^+)\ge q_h(\rho_e)\ge q_h(\rho_r)\ge q_h(\pi^-),
\]
and hence
\[
\Gamma^{\mbox{\tiny IM}}
\le
\Gamma_1
\le
\Gamma_2
\le
\Gamma_3
\le
\Gamma_4
\le
\Gamma^{\mbox{\tiny EX}}.
\]
Using \eqref{eq:app_qr_mu}--\eqref{eq:app_qe_mu} and the net-zero balance, we obtain
\[
q_h^\ast(g)=
\begin{cases}
q_h(\pi^+), & g<\Gamma^{\mbox{\tiny IM}},\\
\underline{q}_r+\underline{q}_e-g,
& g\in[\Gamma^{\mbox{\tiny IM}},\Gamma_1),\\
q_h(\rho_e),
& g\in[\Gamma_1,\Gamma_2),\\
\underline{q}_r+\overline{q}_e-g,
& g\in[\Gamma_2,\Gamma_3),\\
q_h(\rho_r),
& g\in[\Gamma_3,\Gamma_4),\\
\overline{q}_r+\overline{q}_e-g,
& g\in[\Gamma_4,\Gamma^{\mbox{\tiny EX}}],\\
q_h(\pi^-), & g>\Gamma^{\mbox{\tiny EX}},
\end{cases}
\]
\[
q_r^\ast(g)=
\begin{cases}
\underline{q}_r, & g<\Gamma_3,\\
q_h(\rho_r)+g-\overline{q}_e, & g\in[\Gamma_3,\Gamma_4),\\
\overline{q}_r, & g\ge\Gamma_4,
\end{cases}
\]
and
\[
q_e^\ast(g)=
\begin{cases}
\underline{q}_e, & g<\Gamma_1,\\
q_h(\rho_e)-\underline{q}_r+g, & g\in[\Gamma_1,\Gamma_2),\\
\overline{q}_e, & g\ge\Gamma_2.
\end{cases}
\]
Using $w_h=\eta_hq_h$, $w_r=\alpha_rq_r$, and $y=\gamma_eq_e$ gives
\eqref{eq:wh_star_h2}, \eqref{eq:wr_star_h2}, and \eqref{eq:y_star_h2}.

\subsubsection*{4) RO-priority case}

Suppose
\[
\pi^-<\rho_e<\rho_r<\pi^+.
\]
In this case, $\mu$ reaches $\rho_r$ before $\rho_e$, so the RO unit becomes
marginal before the electrolyzer. By the same argument as in the
hydrogen-priority case, the internal thresholds are
\begin{align*}
\Gamma_1' &:= \underline{q}_r+\underline{q}_e-q_h(\rho_r),
&
\Gamma_2' &:= \overline{q}_r+\underline{q}_e-q_h(\rho_r),\\
\Gamma_3' &:= \overline{q}_r+\underline{q}_e-q_h(\rho_e),
&
\Gamma_4' &:= \overline{q}_r+\overline{q}_e-q_h(\rho_e).
\end{align*}
Since $q_h(\mu)$ is nondecreasing,
\[
q_h(\pi^+)\ge q_h(\rho_r)\ge q_h(\rho_e)\ge q_h(\pi^-),
\]
and hence
\[
\Gamma^{\mbox{\tiny IM}}
\le
\Gamma_1'
\le
\Gamma_2'
\le
\Gamma_3'
\le
\Gamma_4'
\le
\Gamma^{\mbox{\tiny EX}}.
\]
Using \eqref{eq:app_qr_mu}--\eqref{eq:app_qe_mu} and the net-zero balance, we obtain
\[
q_h^\ast(g)=
\begin{cases}
q_h(\pi^+), & g<\Gamma^{\mbox{\tiny IM}},\\
\underline{q}_r+\underline{q}_e-g,
& g\in[\Gamma^{\mbox{\tiny IM}},\Gamma_1'),\\
q_h(\rho_r),
& g\in[\Gamma_1',\Gamma_2'),\\
\overline{q}_r+\underline{q}_e-g,
& g\in[\Gamma_2',\Gamma_3'),\\
q_h(\rho_e),
& g\in[\Gamma_3',\Gamma_4'),\\
\overline{q}_r+\overline{q}_e-g,
& g\in[\Gamma_4',\Gamma^{\mbox{\tiny EX}}],\\
q_h(\pi^-), & g>\Gamma^{\mbox{\tiny EX}},
\end{cases}
\]
\[
q_r^\ast(g)=
\begin{cases}
\underline{q}_r, & g<\Gamma_1',\\
q_h(\rho_r)+g-\underline{q}_e, & g\in[\Gamma_1',\Gamma_2'),\\
\overline{q}_r, & g\ge\Gamma_2',
\end{cases}
\]
and
\[
q_e^\ast(g)=
\begin{cases}
\underline{q}_e, & g<\Gamma_3',\\
q_h(\rho_e)-\overline{q}_r+g, & g\in[\Gamma_3',\Gamma_4'),\\
\overline{q}_e, & g\ge\Gamma_4'.
\end{cases}
\]
Using $w_h=\eta_hq_h$, $w_r=\alpha_rq_r$, and $y=\gamma_eq_e$ gives
\eqref{eq:wh_star_water}, \eqref{eq:wr_star_water}, and
\eqref{eq:y_star_water}.

Combining the import, net-zero, and export modes yields
\[
z^\ast(g)=
\begin{cases}
>0, & g<\Gamma^{\mbox{\tiny IM}},\\
=0, & \Gamma^{\mbox{\tiny IM}}\le g\le\Gamma^{\mbox{\tiny EX}},\\
<0, & g>\Gamma^{\mbox{\tiny EX}}.
\end{cases}
\]
This proves the outer mode structure and the closed-form dispatch
expressions in Theorem~\ref{thm:optimal_dispatch}. \IEEEQED

\section{Remarks on Other Orderings and Boundary Cases}
\label{app:other_orders}

Theorem~\ref{thm:optimal_dispatch} focuses on $\pi^-<\rho_r,\rho_e<\pi^+$. Apart from the tie $\rho_r=\rho_e$, the two strict orderings $\rho_r<\rho_e$ and $\rho_r>\rho_e$ give the hydrogen-priority and RO-priority patterns in Section~\ref{sec:OptDis}. If $\pi^-<\rho_r=\rho_e<\pi^+$, the plant is indifferent between allocating the marginal net-zero power increment to RO production or electrolysis. The optimizer may then be nonunique, although the optimal value is unchanged.

For a general ordering, let $i\in\{r,e\}$ and denote its lower and upper power bounds by $\underline q_i$ and $\overline q_i$. The outer-mode solutions are
\begin{equation*}
\begin{aligned}
q_i^{\mbox{\tiny IM}}&\in
\begin{cases}
\{\underline q_i\},&\rho_i<\pi^+,\\
[\underline q_i,\overline q_i],&\rho_i=\pi^+,\\
\{\overline q_i\},&\rho_i>\pi^+,
\end{cases}\\[-1mm]
q_i^{\mbox{\tiny EX}}&\in
\begin{cases}
\{\underline q_i\},&\rho_i<\pi^-,\\
[\underline q_i,\overline q_i],&\rho_i=\pi^-,\\
\{\overline q_i\},&\rho_i>\pi^-.
\end{cases}
\end{aligned}
\end{equation*}
Thus, for strict inequalities, the generalized outer thresholds are
\begin{equation*}
\begin{aligned}
\Gamma_{\mbox{\tiny IM}}^{\rm gen}
&:=q_r^{\mbox{\tiny IM}}+q_e^{\mbox{\tiny IM}}-q_h(\pi^+),\\
\Gamma_{\mbox{\tiny EX}}^{\rm gen}
&:=q_r^{\mbox{\tiny EX}}+q_e^{\mbox{\tiny EX}}-q_h(\pi^-).
\end{aligned}
\end{equation*}
These expressions reduce to the thresholds in Theorem~\ref{thm:optimal_dispatch} under its interior-price assumption.

Within the net-zero mode, a resource with $\rho_i<\pi^-$ remains at $\underline q_i$, whereas one with $\rho_i>\pi^+$ remains at $\overline q_i$. Hence, if only one of $\rho_r$ and $\rho_e$ lies in $(\pi^-,\pi^+)$, only that resource can become marginal and the four internal thresholds reduce to at most two. If neither lies in this interval, no flexible load becomes marginal and the internal net-zero segments degenerate.

The outer and internal thresholds are algebraic quantities and need not be nonnegative. Since $g\in\mathbb R_+$, a negative threshold places the corresponding transition outside the feasible renewable-output domain, so one or more realized modes or segments may disappear. In particular, $\Gamma_{\mbox{\tiny IM}}^{\rm gen}\le0$ leaves no feasible import interval, while $\Gamma_{\mbox{\tiny EX}}^{\rm gen}<0$ places every feasible $g$ in export mode. These cases do not change the threshold form of the policy.

Finally, $\rho_i=\pi^-$ or $\rho_i=\pi^+$ is a singular boundary at which resource $i$ is indifferent at one market margin. The optimizer may then be nonunique or require a tie-breaking convention, but no new qualitative dispatch pattern arises. Therefore, the two cases in Theorem~\ref{thm:optimal_dispatch} exhaust the generic strict interior orderings, while other orderings follow by changing the outer active bounds and degenerating internal thresholds.

\bibliographystyle{IEEEtran}
\bibliography{refs}

\end{document}